\documentclass[aps,pra,twocolumn,showpacs,letterpaper,superscriptaddress]{revtex4-1}

\usepackage[colorlinks=true, citecolor=blue, linkcolor=blue, urlcolor=blue]{hyperref}
\usepackage{graphicx,dcolumn,longtable,epsfig}
\usepackage[usenames]{color}
\usepackage{amssymb}
\usepackage{amsmath}

\usepackage{bm}
\usepackage{footnote}
\usepackage{float}
\usepackage{subfigure}
\usepackage{color}

\usepackage{ulem}
\usepackage[T1]{fontenc}

\newcommand{\be}{\begin{equation}}
\newcommand{\ee}{\end{equation}}

\def\bea{\begin{eqnarray}}
\def\eea{\end{eqnarray}}

\begin{document}
\title{Dipolar bosons in a twisted bilayer geometry}

\author{Chao Zhang}
\affiliation{Department of Physics, Anhui Normal University, Wuhu, Anhui 241000, China}
\author{Zhijie Fan}
\email{zfanac@ustc.edu.cn}
\affiliation{Department of Modern Physics, University of Science and Technology of China, Hefei, Anhui 230026, China}
\affiliation{Hefei National Laboratory, University of Science and Technology of China, Hefei, Anhui 230088, China}
\author{Barbara Capogrosso-Sansone}
\affiliation{Department of Physics, Clark University, Worcester, Massachusetts 01610, USA}
\author{Youjin Deng}
\email{yjdeng@ustc.edu.cn}
\affiliation{Department of Modern Physics, University of Science and Technology of China, Hefei, Anhui 230026, China}
\affiliation{Hefei National Laboratory, University of Science and Technology of China, Hefei, Anhui 230088, China}

\begin{abstract}

In recent years, twisted bilayer systems such as bilayer graphene have attracted a great deal of attention as the twist angle introduces a degree of freedom which can be used to  non-trivially modify system properties. This idea has  been picked up in the cold atom community, first with a theoretical proposal to simulate twisted bilayers in state-dependent optical lattices, and, more recently, with an experimental realization of twisted bilayers with bosonic atoms in two different spin states.
In this manuscript, we theoretically investigate dipolar bosons in a twisted bilayer geometry. The interplay between dipolar interaction and the twist between the layers results in the emergence of quantum states not observed in the absence of twist. We study how system properties vary as we change the twist angle  at fixed distance between the layers and fixed dipolar interaction. We find that at a twist angle $\theta=0.1^{\circ}$, the observed quantum phases are consistent with those seen in the absence of twist angle, i.e. paired superfluid, paired supersolid, and paired solid phases. However, a slight increase in the twist angle to $\theta=0.2^{\circ}$ disrupts these paired  phases in favor of a phase separation between checkerboard solid and superfluid regions. Notably, at a twist angle of $\theta=5.21^{\circ}$, the local occupation number follows the moiré pattern of the underlying moiré bilayers so that a periodic structure of insulating islands is formed. These insulating islands are surrounded by a superfluid. 

\end{abstract}

\maketitle

\section{Introduction}

When two-dimensional periodic structures are overlapped and rotated with respect to each other, new periodic structures, moiré patterns, emerge.  In recent years, condensed matter and material scientists have shown that this mechanism can be used to non-trivially modify properties of two-dimensional systems. At certain twist angles (magic angles), properties of bilayer graphene dramatically change~\cite{NatureCao2018}. What was a weakly-correlated Fermi liquid becomes a strongly-correlated system which supports superconductivity,  quantized anomalous Hall states, and more~\cite{Andrei:2020tp}. The quantum anomalous Hall effect has also been observed with moiré hetero-bilayers~\cite{doi:10.1126/science.aay5533,Li:2021wr}. 
The field of moiré bilayers remains active~\cite{PhysRevLett.128.026402,Zhao:2024uw,PhysRevX.14.011004,Zhao:2024aa}.

Moiré bilayers have also been engineered with ultracold gases. The unprecedented level of control in these setups provides an ideal setting to simulate condensed matter systems in a highly controlled environment~\cite{Bloch:2012ty,  Schafer2020}. A few years ago, a theoretical proposal to simulate twisted bilayers with cold atoms trapped in state-dependent optical lattices was put forward~\cite{PhysRevA.100.053604}. In a recent experiment~\cite{Meng:2023wk}, twisted bilayers at the magic angle $\theta=5.21^{\circ}$ have been artificially realized using bosonic atoms in two different spin states trapped in spin-dependent square optical lattices.
This introduces yet another facet to the cold atoms toolbox -- an ideal platform to investigate the intricate interplay between lattice geometry, interactions, and quantum fluctuations. The tunability of these systems, achieved by adjusting the twist angle between layers, provides a unique opportunity to manipulate quantum many-body systems at the microscopic level and explore the emergence of exotic states of matter. These studies are driven not only  by a fundamental curiosity about the behavior of matter in these artificial structures but also by the potential applications in quantum simulation and quantum information processing.

Lattice bosons interacting via long-range, anisotropic dipolar interaction can stabilize various solid and supersolid phases~\cite{PhysRevA.105.063302, PhysRevA.103.043333, Zhang_2015, PhysRevA.106.063313, chomaz_dipolar_2022}. Recently, various solid phases  have been experimentally observed with cold magnetic atoms~\cite{Nature2023Su}. Dipolar bosons in layered geometries can also induce pairing leading to paired solid, paired superfluid, and paired supersolid phases~\cite{Capogrosso-Sansone:2011tf,Safavi-Naini_2013,PhysRevA.90.043604}. In this work, we consider dipolar lattice bosons trapped in a twisted bilayer geometry. Layers are placed at a fixed  distance $d_z$. Particles cannot hop between layers and a fixed dipolar interaction provides coupling between particles on different layers. We are interested in investigating the effects on the phase diagram of the bilayer geometry when a finite, small twist angle  between the layers is introduced. Our findings can be summarized as follows. At small enough twist angle $\theta \sim 0.1^{\circ}$, the quantum phases and phase transitions remain consistent with those observed at zero twist angle~\cite{Safavi-Naini_2013}. That is, paired superfluid, paired checkerboard solid, and paired supersolid phases are stabilized.  However, a slight increase in the twist angle to $\theta = 0.2^{\circ}$ disrupts the paired superfluid and paired supersolid phases in favor of  phase separation between the checkerboard solid and superfluid phases. At the magic angle $\theta = 5.21^{\circ}$, the local occupation number follows the moiré pattern of the underlying moiré bilayers so that a periodic structure of insulating islands is formed. These insulating islands are surrounded by a superfluid. 

This paper is organized as follows: In Sec.~\ref{sec:sec2}, we introduce the Hamiltonian of the system. In Sec.~\ref{sec:sec3}, we discuss various phases, the corresponding order parameters and correlators. In Sec.~\ref{sec:sec4}, we present the results of the system for three small fixed values of twisted angles $0.1^{\circ}, 0.2^{\circ}$, and magic angle $5.21^{\circ}$; we outline our conclusions in Sec.~\ref{sec:sec5}.

\begin{figure}
\includegraphics[width=0.45\textwidth]{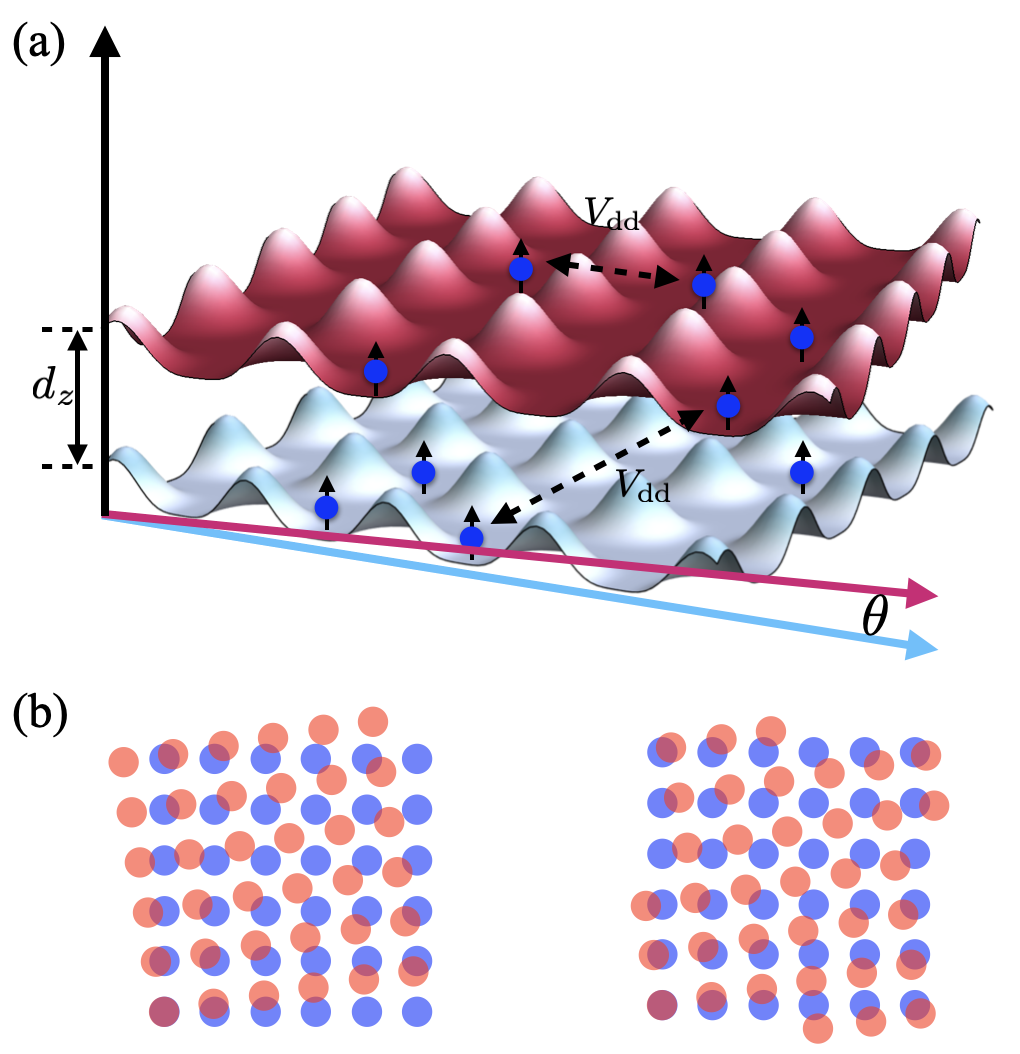}
\caption{(a) Schematic representation of the twisted bilayer system. Dipoles are trapped in a bilayer lattice and are aligned parallel to each other (and perpendicular to the lattice layers) along the direction of polarization. With this geometry, the intra-layer interaction is purely repulsive while dipoles sitting on top of each other in different layers attract. The distance between the two layers is $d_z$. The upper layer is twisted by an angle $\theta$ with respect to the lower layer. (b) Schematic illustration of the boundary conditions.
} 
\label{setup}
\end{figure}

\section{Hamiltonian}
\label{sec:sec2}

In the single band approximation, the system can be described by the Hamiltonian~\cite{Safavi-Naini_2013}:

\begin{align}
H=-t \sum_{\langle i,j \rangle, \alpha} a_{i\alpha}^{\dagger} a_{j\alpha} +\frac{1}{2} \sum_{i \alpha; j\beta} V_{i \alpha; j\beta} n_{i,\alpha} n_{j,\beta} - \sum_{i,\alpha} \mu_{\alpha} n_{i,\alpha}.  
\label{Eq1}
\end{align}
Here, $\alpha, \beta=1,2$ and $i,j$ label the layers and the lattice sites in each layer respectively,  $a_{i,\alpha} (a_{i,\alpha}^{\dagger})$ are the bosonic annihilation (creation) operators obeying the hard-core constraint $a_{i,\alpha}^{\dagger 2}=0$, and $n_{i,\alpha}=a_{i,\alpha}^{\dagger}a_{i,\alpha}$ is the number operator for particles. The brackets $\langle \cdots \rangle$ denote summation over nearest neighbors only. The first term in equation~\ref{Eq1} describes the kinetic energy with in-plane hopping rate $t$. The second term is the dipole-dipole interaction given by $V_{i \alpha; j \beta}=C_{\text{dd}}(1-3 \text{cos}^2 \gamma_{i_{\alpha}; j_{\beta}}) /4 \pi |r_{i_{\alpha}; j_{\beta}}|^3$, where $\gamma_{i_{\alpha}; j_{\beta}}$ characterizes the angle between
the direction of the polarization and the relative position
of the two particles given by $r_{i_{\alpha}; j_{\beta}}$, and $C_{\text{dd}}=d^2/\epsilon_0(C_{\text{dd}}=\mu_0 d^2)$ for electric (magnetic) dipoles of strength $d$. The repulsive  nearest neighbor intra-layer interaction is $V_{\text{dd}}=C_{\text{dd}}/(4\pi a^3)$, with $a$ the in-plane lattice constant. The attractive  nearest neighbor inter-layer interaction is $V_{\text{dd}}^{\perp}=-2 C_{\text{dd}}/4\pi d_{z}^3$, with $d_z$ the inter-layer distance. The quantity $\mu_{\alpha}$ is the chemical potential which sets the number of particles in each layer. Here, we fix $\mu_1=\mu_2$, i. e. $N_1=N_2$. The twist angle between these two layers is $\theta$.

In the following, we present unbiased results based on path-integral quantum Monte Carlo using the two-worm algorithm~\cite{Soyler_2009, Lingua_2018}. 
We have performed the simulations with system size $L\times L=N_{site}$ for each layer, with $L=20, 22, 33$ (we choose the lattice constant $a$ to be our unit of length). The filling factor is $n=N/N_{site}$ with $N=N_1=N_2$ the particle number on each layer. The inverse temperature $\beta$ is set to $\beta=L$. As for boundary conditions, each layer is {\it separately} mapped into a torus. As a result, lattice sites on the twisted layer which ``fall outside'' the region covered by the untwisted layer are considered within the untwisted region, and particle position and relative particle distances entering the hamiltonian are calculated accordingly. In other words, we set the simulation box in each direction to be $(1/2,L+1/2]$, see sketch in Fig.~\ref{setup}(b). With this approach, we maximize  homogeneity of the system and reduce boundary effects so that our results highlight the effect of the twisted geometry on system properties.  Due to computational limitations, our simulations are restricted to a system size of $L=33$. Due to this limitation, with open boundary conditions, we would expect boundary effects to be conflated with the effects of the twisted geometry. By using the torus-mapped boundary, we can avoid this issue and isolate the impact of the twisted geometry.

\section{Order parameters and correlators}
\label{sec:sec3}

To discern superfluid phase (SF) from paired superfluid phase (PSF), we measure the superfluid stiffness $\rho_s$ for each layer and the superfluid stiffness for pairs $\rho_{\text{PSF}}$.
The superfluid stiffness is $\rho_s=\langle \mathbf{W}^2 \rangle /dL^{d-2}\beta$~\cite{Ceperley:1989hb}, where $\langle\mathbf{W}^2\rangle=\sum_{i=1}^d\langle W_i^2\rangle$ represents the expectation value of the winding number squared, $d$ is the dimension of the system (in our case, $d=2$), $L$ is the linear system size, and $\beta$ is the inverse temperature. The superfluid stiffness for pairs is directly associated with a pair condensate and is given by $\rho_{\text{PSF}}= \langle \mathbf{W}_{+}^2 \rangle /d L^{d-2} \beta$, where $\mathbf{W}_{+} = \mathbf{W}_1 + \mathbf{W}_2$ represents the sum of winding numbers in layers 1 and 2
and can be calculated within the two-worm path-integral quantum Monte Carlo. We note that, due to pairing across the layers, in the PSF phase the fluctuation of difference in winding numbers is zero, $\langle (\mathbf{W}_1 - \mathbf{W}_2)^2 \rangle=0$.  Diagonal long-range order, instead, corresponds to finite structure factor, 
 $  S(\mathbf{k}) = \sum_{\mathbf{r},\mathbf{r'}} \exp[i \mathbf{k} (\mathbf{r}-\mathbf{r'})\langle n_r n_{r'} \rangle]/N, 
$ with $N$ the particle number. $\mathbf{k}$ is the reciprocal lattice vector. Here, $\mathbf{k}=(\pi,\pi)$ to identify a checkerboard (CB) density pattern.

In addition, we also measure two-point  correlation functions for each species $f_\alpha (r_{i_{\alpha};j_{\alpha}})\propto\langle a_{i\alpha} a^\dag_{j\alpha} \rangle$, where the vector $r_{i_{\alpha};j_{\alpha}}=(x_{i_{\alpha}}-x_{j_{\alpha}},y_{i_{\alpha}}-y_{j_{\alpha}})=(x_{ij},y_{ij})$ is the relative position of the annihilation and creation operators at layer $\alpha$ (to ease the notation, we dropped the label $\alpha$ in the coordinates representation). Notice that, here, to calculate relative distances between annihilation and creation operators on each layer, we consider the relative position vector to originate at the site of $a^\dag$   so that the maximum distance in $x$ and $y$ directions is $L-1$.
Here, $\langle\rangle$ denotes a quantum and thermal average. $f_\alpha (r_{i_{\alpha};j_{\alpha}})$ is normalized so that its maximum value is 1. This correlator is expected to be long-ranged in a SF phase and short-ranged  in an insulating phase. 

Finally, we also produce \textit{ density maps} and \textit{ condensate maps}. The first is a map of the average of the local occupation number $\langle n_{i_\alpha}\rangle$ of a single typical Monte Carlo configuration. The latter is a map of the probability distribution for lattice sites to be visited by $ a_{i\alpha}$ and $a^\dag_{j\alpha}$ when their distance is larger than a chosen cutoff $R_c$. This map informs us on the lattice regions where off-diagonal long-range order exists.

\section{Results and discussion}
\label{sec:sec4}

Let us start with a description of the phase diagram in the absence of twist and at fixed $d_z=0.36$. The phase diagram is comprised of pair checkerboard solid (PCB), pair supersolid (PSS), paired superfluid (PSF), and, at lower dipolar interaction, independent superfluids (2SF)~\cite{ Safavi-Naini_2013}. 
In the PCB phase, reached at half filling and large enough dipolar interaction,  atoms across the layers are strongly
paired due to attractive interlayer interactions, i.e., particles sit on top of each other. The system can be thus envisioned as a solid of pairs. Upon doping the PCB solid with particles or holes, the PSS phase is stabilized. In this phase, CB density-density correlations coexist with a finite superfluid stiffness associated with pairs, i.e. $\rho_{\text{PSF}}\ne 0$ while $\langle (\mathbf{W}_1 - \mathbf{W}_2)^2 \rangle =0$. For large enough doping, density-density correlation are lost (via a 2+1 Ising transition), and the system becomes a PSF. At lower values of dipolar interaction, a 2SF phase can be stabilized.

In the following, we fix the distance between the two layers to $d_z=0.36$, and the intra-layer dipolar interaction to $V_{\text{dd}}/t=0.26$. We denote the repulsive (attractive) nearest neighbor intra-layer (inter-layer)
interaction by $V_{\text{dd}}=C_{\text{dd}}/(4\pi a^3)(V_{\text{dd}}^{\perp}=-2C_{\text{dd}}/4 \pi d_z^3)$, with $a$ the in-plane lattice constant. These choices are made to compare with existing results on bilayer systems with no twist between the layers~\cite{Safavi-Naini_2013}. We consider twist angles $\theta=0.1^{\circ}$, $0.2^{\circ}$, and magic angle $5.21^{\circ}$.

\subsection{Case of $\theta=0.1^{\circ}$} 

\begin{figure}
\includegraphics[width=0.48\textwidth]{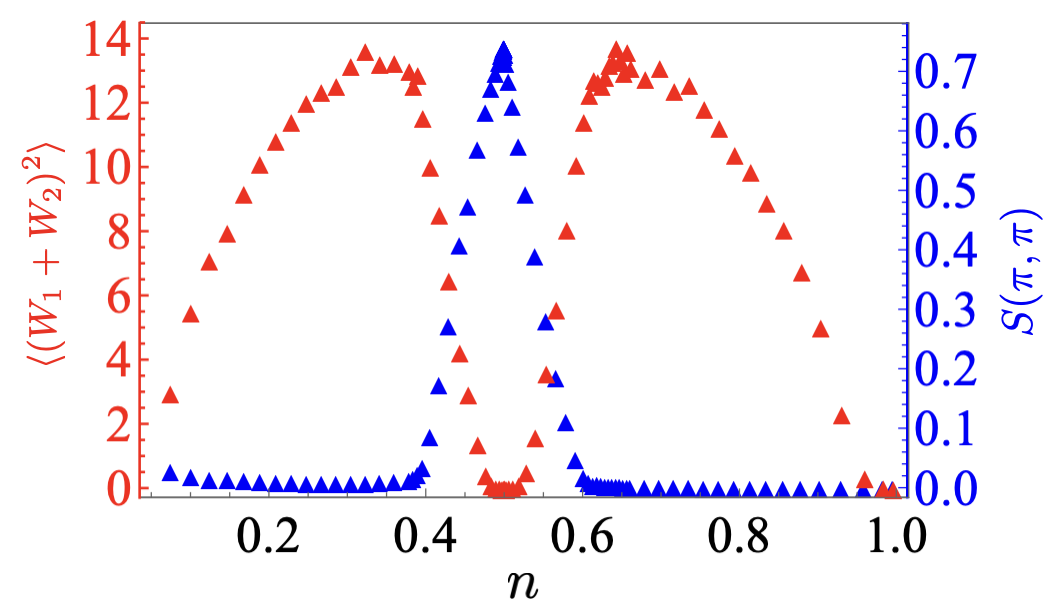}
\centering
\caption{The expectation value of the square of the sum of winding numbers $\langle(\mathbf{W}_1 + \mathbf{W}_2)^2 \rangle$ and the structure factor $S(\pi, \pi)$ as functions of density $n$. $V_{\text{dd}}/t=0.26$,  $d_z=0.36$, $L=22$,  twist angle $\theta=0.1$. $ \langle(\mathbf{W}_1 - \mathbf{W}_2)^2\rangle $ stays zero for the whole density range.
} 
\label{FIG2}
\end{figure}

In our initial exploration, we focus on a modest twist angle, $\theta=0.1^{\circ}$. At this twist angle, we do not see any qualitative difference with the case of $\theta=0^{\circ}$.  This phase diagram remains largely unchanged at twist angle $\theta=0.1^{\circ}$. As an example, in the main plot of figure~\ref{FIG2}, we plot the expectation value of the sum of winding numbers $\langle \mathbf{W}_{+}^2 \rangle= \langle(\mathbf{W}_1 + \mathbf{W}_2)^2\rangle$ and the structure factor $S(\pi, \pi)$ as a function of density $n$ at fixed intra-layer dipolar interaction $V_{\text{dd}}/t=0.26$ and system size  $L=22$. 
From this plot, it is clear that at $n=0.5$ CB density correlations are present while $\rho_{\text{PSF}}$ (and $\rho_s$) is zero. Away from half filling, there exists a range of densities for which both  $S(\pi, \pi)$ and $\langle \mathbf{W}_{+}^2 \rangle$ are finite while $\langle \mathbf{W}_{-}^2 \rangle=0$. For these densities, the system is in a PSS phase. Eventually, CB density correlations disappear at $n \sim 0.41$, in agreement with the case of $\theta=0^{\circ}$, and the system becomes a PSF.

\subsection{Case of $\theta=0.2^{\circ}$}

\begin{figure*}
\includegraphics[width=1.02\textwidth]{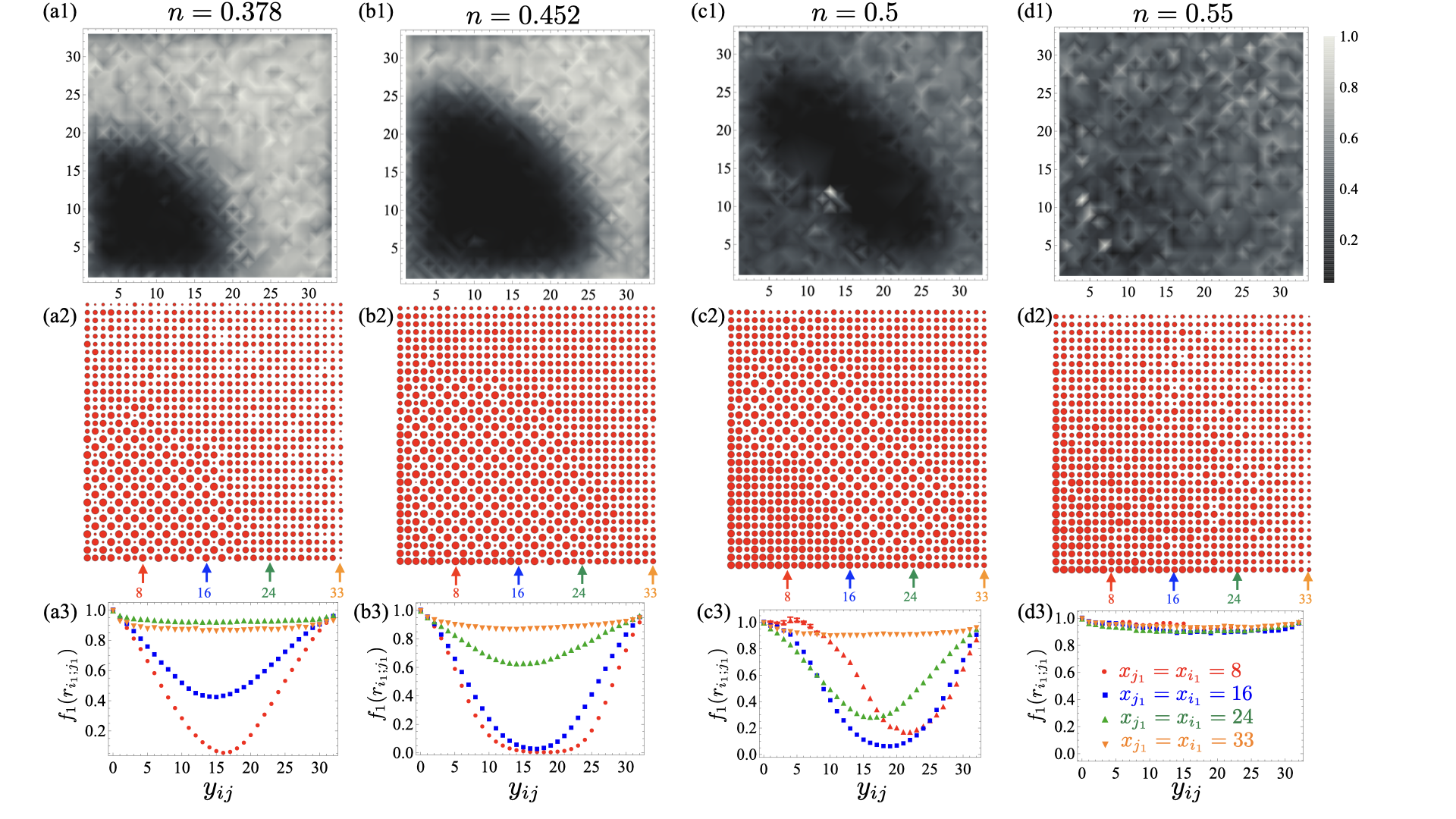}
\centering
\caption{Condensate map (a1), density map  (a2) of the top layer (maps for the bottom layer look very similar) and two-point correlation function $f_1 (r_{i_{1};j_{1}})$ (a3) ($f_2 (r_{i_{2};j_{2}})$ looks very similar) with $x_{j_{1}}=x_{i_{1}}=8,16,24,33$ (red dots, blue squares, green upper-triangles, and orange down-triangles) as a function of $y_{ij}$, for twist angle $\theta=0.2^{\circ}$ filling factor $n=0.378$ and system size $L=33$. Same plots for filling factor $n=0.452$ (b1-b3),  $n=0.5$ (c1-c3), and $n=0.55$ (d1-d3).
} 
\label{FIG3}
\end{figure*}

A marginal increase in the twist angle to $\theta=0.2^{\circ}$ introduces notable changes to the system properties. Our simulation results reveal that PSS is destabilized in favor of phase separation, where CB solid and SF orders are both present but in different regions of the lattice, while PSF is destabilized in favor of independent SF. The second row of Figure~\ref{FIG3} shows the density map at fixed filling factors $n=0.378$ (a2), $n=0.452$ (b2), $n=0.5$ (c2), $n=0.55$ (d2) for system size $L=33$. Here, each circle corresponds to a single lattice site, and its radius is proportional to the local density. The map refers to the top layer. The density map for the bottom layer is very similar. We notice that at $n=0.378$ in the bottom-left region of the lattice the system is  a CB.  In the remaining region of the lattice, instead, we observe a roughly uniform density. In this region, off-diagonal long-range order is present but pairing is destroyed since both $\langle \mathbf{W}_{+}^2 \rangle$ and $\langle \mathbf{W}_{-}^2 \rangle$ are finite. The $x$ and $y$ axis separate these two phases and as a result CB order is less pronounced here. In figure~\ref{FIG3} (a3), we plot the two-point correlation function  
$f_1 (r_{i_{1};j_{1}})$ ($f_2 (r_{i_{2};j_{2}})$ looks very similar) with $x_{j_{1}}=x_{i_{1}}=8,16,24,33$ as a function of $y_{ij}$. Along the cut $x=24,33$ the correlator decays to a finite constant $\sim0.9$ at the largest distance on the torus $L/2+1$, signaling the superfluid nature of the system in this region of space. At $x=16$, we observe a significant decay of the correlator to $\sim0.42$. This is because along this cut part of the system is in a CB phase. The CB region does not contribute to a finite correlator at the larger distances on the torus. Indeed, in this region of space, the correlator decays to zero within a few lattice steps. We also notice that the correlator stays finite at the largest distance $L/2+1$ because along this cut the SF region extends for a distance $\sim L/2$. Finally, At $x=8$, the correlator decays to zero at the largest distance. This is because along this cut, the SF region extend for a distance $<L/2$.

To further verify our findings, in figure~\ref{FIG3} (a1) we plot the condensate map corresponding to the same parameters. Here,  the region of space corresponding to a CB arrangement of particle density is dark, corresponding to negligible probability within the region 
of finding creation and annihilation operators at distance further than $R_c=5$. This means that no off-diagonal long-range order is present in this region of space.  In the region of space where we observe roughly uniform density in figure~\ref{FIG3} (a2), instead, there exists a finite probability to find creation and annihilation operators at distance further than $R_c$ (light region).

At $n=0.452$, only quantitative changes are observed, figure~\ref{FIG3}(b1-b3).  
In figure~\ref{FIG3} (b2), we observe 
 a noticeable expansion of the CB region. We notice that, at this density, the CB order is not present  along the $x$- and $y$-axis. In figure~\ref{FIG3} (b3), we plot the two-point correlation function  
$f_1 (r_{i_{1};j_{1}})$  with $x_{j_{1}}=x_{i_{1}}=8,16,24,33$ as a function of $y_{ij}$. Along the cut $x=33$ the correlator decays to a  finite value ($\sim0.9$) signaling the superfluid nature of the system in this region of space. At $x=24$, the correlator decays to a lesser value of $\sim 0.6$ because along this cut lies the edge of the CB region which partially suppress off-diagonal long-range order at larger distances. At $x=8, 16$,  the correlator decays to zero at the largest distances. This is because along this cut, the SF region only extends for distances of 7-10 sites. All these findings are also reflected in the condensate map of figure~\ref{FIG3}(b1), confirming that even this small twist angle destroys supersolidity.
 
As the density is further increased, the region corresponding to CB solid starts shrinking from the bottom left corner. Noticeably, this remains the case also at half filling, see figure~\ref{FIG3}(c1-c3). In figure~\ref{FIG3}(c3), we notice that the correlators stay finite along all the cuts, though they do decay to different values  depending on the thickness of the CB region along the corresponding cut.
Eventually, CB order  disappears altogether at large enough density ($n \sim 0.53$) and the system is a SF, see figure~\ref{FIG3}(d1-d3) for filling factor $n=0.55$.

\subsection{Case $\theta=5.21^{\circ}$}

\begin{figure}
\includegraphics[width=0.45\textwidth]{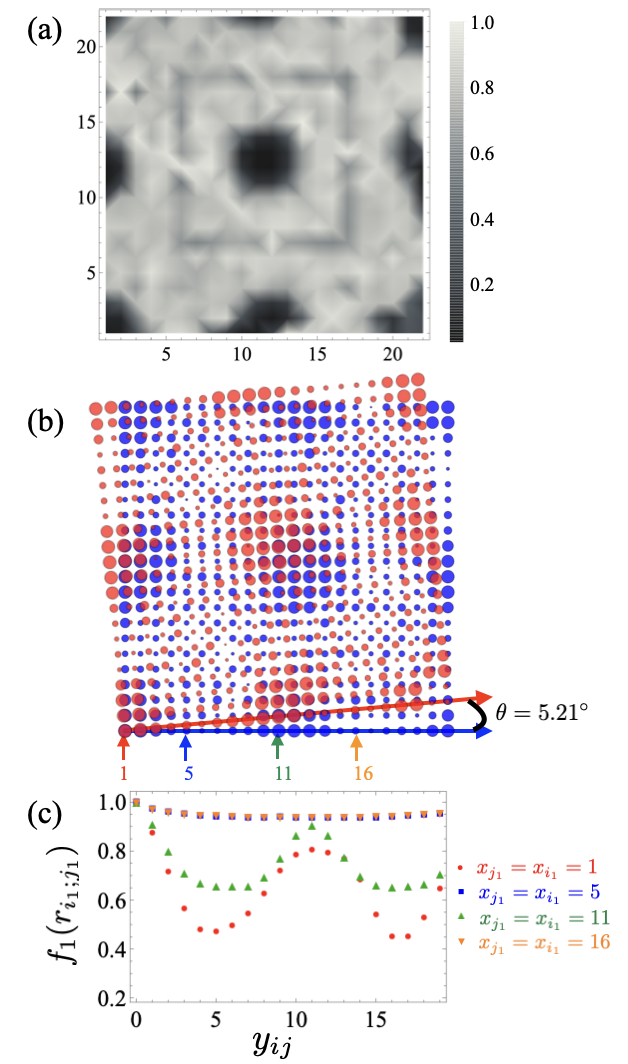}
\centering
\caption{
(a) Condensate map, (b) density map of the top (red) and bottom (blue) layers at filling factor $n=0.652$ for system size $L=22$. (c) two-point correlation function  $f_1 (r_{i_{1};j_{1}})$ ($f_2 (r_{i_{2};j_{2}})$ looks very similar) with $x_{j_{1}}=x_{i_{1}}=1,5,11,16$ (red dots, blue squares, green upper-triangles, and orange down-triangles) as a function of $y_{ij}$.
}
\label{FIG4}
\end{figure}

In order to attain commensurate superlattices in a square lattice,  twist angles need to  satisfy the condition $\theta=2\text{arctan}(m/n)$, where $m$ and $n$ stand as coprime natural numbers~\cite{NatureWang2020}. In this work, we have chosen  $\theta = 5.21^{\circ}\sim 2 \text{arctan}(1/22)$. As a result supercells manifest a periodicity denoted by $\lambda_{mo} \sim 11$. The density maps in figure~\ref{FIG4}(b) for $n=0.652$ reveal the presence of periodically-spaced Mott insulator (MI) islands with the expected periodicity $\lambda_{\rm{mo}} \sim 11$. These islands are surrounded by SF regions. Here, red (blue) circles refer to the top (bottom) layer. This moiré pattern has been observed experimentally with two-component systems utilizing Bose-Einstein condensates loaded into spin-dependent optical lattices. 

The periodic arrangement of MI islands is also evident in figure~\ref{FIG4}(c) where we plot the two-point correlation function  
$f_1 (r_{i_{1};j_{1}})$ ($f_2 (r_{i_{2};j_{2}})$ looks very similar) with $x_{j_{1}}=x_{i_{1}}=1,5,11,16$ as a function of $y_{ij}$. Along the cut $x=5,16$ the correlator decays to a finite constant value  ($\sim0.95$) signaling the superfluid nature of the system along this cut.  At $x=1,11$, we notice the periodic nature of the correlator with periodicity equals to $\lambda_{\rm{mo}}=11$, the expected periodicty of the observed moiré pattern. For these correlators, the first minimum is observed at $y_{ij}\sim 5$ corresponding to the width of the MI islands. These islands are insulating and therefore do not contribute to the finiteness of the correlator at these distances. Similarly, the next minimum appears at  $y_{ij}\sim 5+\lambda_{\rm{mo}}$, corresponding to the next MI island. We also notice that along the cut at $x=11$ the value of the minimum is larger than for $x=1$. This is because  this cut is not centered within the moiré pattern, being situated one lattice spacing away from the central position which in this case is $x=12$. 

Finally, in figure~\ref{FIG4} (a) we plot the condensate map corresponding to the same parameters. Here, we clearly see the periodically-spaced  MI islands to appear as dark, corresponding to lack of off-diagonal long-range order in these regions of space.

\section{Conclusion}
\label{sec:sec5}
In conclusion, our investigation into dipolar bosons confined within a twisted bilayer geometry has uncovered a wealth of intriguing phenomena. We have elucidated the profound impact of twisted geometries on the quantum phases exhibited by the system. At a small twist angle of $\theta=0.1^{\circ}$, the system behavior remains consistent with that of the untwisted case. However, a slight increase in the twist angle to $\theta=0.2^{\circ}$ induces significant changes, disrupting paired  phases and leading to the emergence of phase separation phenomena. This observation highlights the sensitivity of the system to slight alterations in the twisted geometry. At larger twist angle  $\theta=5.21^{\circ}$, we have observed that particle density follows the pattern of the underlying  moiré bilayer resulting in the emergence of periodic Mott insulator islands surrounded by superfluid regions, a phenomenon not observed in untwisted systems. This discovery highlights the novel and unexpected behaviors that can arise from the interplay between twisted geometries and dipolar interactions.

Our findings offer valuable insights into the interplay between geometry, interactions, and quantum fluctuations in correlated bosonic systems with potential implications for cold-atom experiments. By harnessing the precision and control afforded by cold-atom setups, the phenomena observed theoretically can be further explored and validated, advancing our understanding of complex quantum systems.

\begin{acknowledgments}
CZ, ZF, and YD acknowledge the support by the National Natural Science Foundation of China (NSFC) under Grant Nos. 12204173, 12275263, 12275002, and the Innovation Program for Quantum Science and Technology (under Grant No. 2021ZD0301900). CZ also acknowledges the support by the University Annual Scientific Research Plan of Anhui Province under Grant No 2022AH010013. YD also acknowledges the support by the Natural Science Foundation of Fujian province of China (under Grant No. 2023J02032). The computing for this project was performed at the cluster at Clark University.
\end{acknowledgments}

\bibliography{twisted}

\end{document}